\documentclass[12pt,a4paper]{article} 
\usepackage{a4wide}
\usepackage{epsfig}
\usepackage{graphics}
\usepackage{subfigure}

\title{Periodic Monopoles}
 \author{R.\ S.\ Ward\footnote{email: richard.ward@durham.ac.uk}
 \bigskip
\\Department of Mathematical Sciences,  \\ University of
Durham, \\Durham DH1 3LE}

\newcommand{\tr}{\mathop{\rm tr}\nolimits}

\newcommand{\pa}{\partial}
\newcommand{\ii}{{\rm i}}

\newcommand{\RR}{{\bf R}}

\begin{document}

\maketitle
\abstract{
\noindent This paper deals with static BPS monopoles in three dimensions
which are periodic either in one direction (monopole chains) or two
directions (monopole sheets).  The Nahm construction of the simplest
monopole chain
is implemented numerically, and the resulting family of solutions described.
For monopole sheets, the Nahm transform in the U(1) case is computed
explicitly, and this leads to a description of the SU(2) monopole sheet
which arises as a deformation of the embedded U(1) solution.
}

\vskip 1truein
\noindent PACS 11.27.+d, 11.10.Lm, 11.15.-q

\newpage


\section{Introduction}

In recent years, there has been interest in periodic BPS monopoles, namely
solutions of the Bogomolny equations on $\RR^3$ which are periodic either
in one direction (monopole chains) or two directions (monopole sheets).
This has arisen partly because of the interpretation and application of
such solutions in the theory
of D-branes.  For monopole chains, the details of the Nahm transform
have been fully explored, and there are some partial existence results
\cite{CK01}; but for monopole sheets, much less is known \cite{L99}.
In neither case are there any explicit solutions.  The main purpose of this
Letter is to review what is known about the simplest (unit charge) 
monopole chains and monopole sheets, and to describe their appearance.

In the chain case, we implement the Nahm construction
numerically, to obtain a one-parameter family of 1-monopole chains;
the parameter is the ratio between the monopole size and the period.
In the sheet case, there is a homogeneous U(1) monopole sheet solution;
we demonstrate that this is ``self-reciprocal'' under the Nahm transform,
and describe the appearance of the SU(2) monopole sheet which arises
as a deformation of this abelian solution.

The fields we deal with are solutions of the Bogomolny equations
\begin{equation} \label{Bog}
  D_j \Phi = -\frac{1}{2}\varepsilon_{jkl} F_{kl}
\end{equation}
on $\RR^3$. Here the coordinates are $x^j=(x,y,z)$, the gauge field is
$F_{jk}=\pa_j A_k - \pa_k A_j +[A_j, A_k]$, and 
$D_j \Phi=\pa_j \Phi +[A_j, \Phi]$. We take the gauge group to be SU(2);
except that in the section on monopole sheets, we start with U(1) fields.
The norm-squared of the Higgs field $\Phi$ is defined by
$|\Phi|^2=-\frac{1}{2}\tr(\Phi^2)$, and the energy density is
\begin{equation} \label{Enden}
  {\cal E} = -\frac{1}{2}\tr\left[(D_j\Phi)^2
          + \frac{1}{2}(F_{jk})^2 \right].
\end{equation}
If (\ref{Bog}) is satisfied, then ${\cal E}=\nabla^2|\Phi|^2$,
where $\nabla^2$ is the Laplacian on $\RR^3$.


\section{Monopole Chains}

In this section, we are interested in BPS monopoles on $\RR^2\times S^1$;
specifically, monopoles which are periodic in $z$ with period $2\pi$.
Let us begin with some general remarks. In the case of periodic instantons
(calorons), one may proceed by taking a finite chain of instantons
($m$ instantons strung along a line in $\RR^4$ with equal spacing), and
letting the number $m$ tend to infinity --- indeed, the first example
of a caloron solution was constructed in this way \cite{HS78}.  For monopoles,
there is a solution representing a string of $m$ monopoles \cite{ES89}: one
can write down
its Nahm data explicitly in terms of the $m$-dimensional irreducible
representation of $su(2)$. But this has no limit as $m\to\infty$,
so one does not get an infinite monopole chain in this way.

There is another way to understand why one expects something to go wrong
in the $m\to\infty$ limit \cite{CK01,L99}.  In the
asymptotic region $\rho^2=x^2+y^2\to\infty$, the Higgs field $\Phi$ of
a chain of single SU(2) monopoles will behave like a chain of U(1) Dirac
monopoles, for which the Higgs field, by linear superposition, is
$\phi = -\frac{1}{2}\sum_{p\in{\bf Z}}
         \left[\rho^2+(z-2\pi p)^2\right]^{-1/2}$.
But this series diverges: the $m$-chain (which corresponds to a finite series)
has no limit as $m\to\infty$. One may, instead, define a
chain of Dirac monopoles by subtracting an infinite constant, to obtain
\begin{equation} \label{Dirac2}
   \phi = \alpha - \frac{1}{2r} - \frac{1}{2} \sum_{p\neq0}
         \left[ \frac{1}{\sqrt{\rho^2+(z-2\pi p)^2}}
         -\frac{1}{2\pi|p|} \right],
\end{equation}
where $\alpha$ is a constant. This field is smooth, except at the locations
$\rho=0$, $z\in2\pi{\bf Z}$ of the monopoles, and
has the asymptotic behaviour $\phi\sim(\log\rho)/(2\pi)$ for large $\rho$.

This U(1) example motivates the boundary conditions for the non-abelian case
\cite{CK01}. In particular, we require that
\begin{equation} \label{chainBC}
  |\Phi| \sim \frac{N}{2\pi}\log\rho, \quad |D\Phi|=O(1/\rho)
\end{equation}
as $\rho\to\infty$, where $N$ is a positive integer. In fact, $N$ is a
topological invariant: the eigenvector of $\Phi$ associated with its
positive eigenvalue defines a line bundle over the 2-torus $\rho=c\gg1$,
and the first Chern number of this line bundle is $N$.  A smooth solution
of (\ref{Bog}) satisfying the boundary condition (\ref{chainBC}) may be
thought of as an infinite chain of $N$-monopoles.

Through the Nahm transform \cite{CK01}, such monopole chains correspond to
solutions of the U($N$) Hitchin equations on the cylinder $\RR\times S^1$,
with appropriate boundary conditions.
Let us concentrate on the $N=1$ case, and describe the Nahm
construction of the $N=1$ monopole chain.  

Write $s=r+\ii t$, where $r\in{\RR}$ and $t\in[0,1)$ are coordinates on the
cylinder.
Let $\Delta$ be the first-order differential operator
\begin{equation} \label{chainDelta}
  \Delta = \left[\begin{array}{cc}
            2\pa_{\bar s}-z & P(s) \\
            \overline{P(s)} & 2\pa_s+z
         \end{array}\right],
\end{equation}
where $P(s)=C\cosh(2\pi s)-(x+\ii y)$, with $C$ being
a positive constant. For each spatial point $x^j=(x,y,z)$,
the $L^2$ kernel of this operator is 2-dimensional. So there exists a
$2\times2$ matrix $\Psi(t,r;\,x^j)$ satisfying
\begin{equation} \label{chainBVP}
  \Delta\Psi=0, \quad
    \int_{-\infty}^{\infty}\int_0^1 \Psi^{\dagger}\Psi\,dt\,dr=I,
\end{equation}
where $I$ denotes the $2\times2$ identity matrix.  Then
\begin{equation} \label{chainField}
  \Phi= \ii\int_{-\infty}^{\infty}\int_0^1 r\Psi^{\dagger}\Psi\,dt\,dr,
  \quad
  A_j =\int_{-\infty}^{\infty}\int_0^1 \Psi^{\dagger}
                            \frac{\pa}{\pa x_j}\Psi\,dt\,dr
\end{equation}
defines a 1-monopole chain satisfying (\ref{Bog}) and (\ref{chainBC}).

The explicit solution of the boundary-value problem (\ref{chainBVP}) is not
known. Part of the difficulty is the lack of symmetry --- both the finite
and the infinite monopole chains seem to have only a $D_2$ symmetry,
corresponding to rotations by $180^\circ$ about each of the $x$, $y$ and $z$
axes. This is quite unlike the situation for the $N=1$ instanton chain,
where one has O(3) symmetry, and an explicit caloron solution \cite{HS78}.
So to see what the monopole chain looks like, one has to solve (\ref{chainBVP})
either approximately or numerically.

This solution contains only one parameter, namely $C$. All the other moduli
can (and have) been removed by translations and rotations of the $x^j$.
(Of course, there will be far more parameters when $N>1$.)
From (\ref{chainDelta}), one might guess on dimensional grounds that
$C$ determines the monopole size, and this is indeed the case; or rather,
since the length-scale is already fixed by the period of $z$, the parameter
$C$ corresponds to the dimensionless ratio between the
monopole size and the $z$-period.

If $0<C\ll 1$, then one would expect to obtain a chain of small monopoles
located at the points $\rho=0$, $z\in2\pi{\bf Z}$ along the $z$-axis; in fact,
like the Dirac chain (\ref{Dirac2}) but with the singularities smoothed out.
A numerical implementation of the Nahm construction produces results that are
consistent with this interpretation.  The more interesting case is $C\gg1$:
namely, what happens when the monopole size becomes greater than the
$z$-period?  It is this question that we shall concentrate on here.

Once again, it is worth contrasting with the caloron case.
The large-size limit of a 1-caloron is in fact a 1-monopole \cite{R79};
but for $N\geq2$, the large-size limit may be an $N$-monopole, or may not
exist at all \cite{W04}.

So let us look for approximate solutions of
\begin{equation} \label{chainDeltaEqn}
  \Delta \left[\begin{array}{c} g\\f\end{array}\right]
  = \left[\begin{array}{c} 
         2g_{\bar s}-zg+Pf\\
         2f_s+Zf+\overline{P}g
         \end{array}\right] = 0
\end{equation}
when $C\gg1$. Clearly the functions $g$ and $f$ will have to be close to zero,
except near the zeros $\pm s_0$ of the function $P(s)$.  In other words, $g$ and
$f$ are supported near the two points
$s=\pm s_0=\pm[\cosh^{-1}(\zeta/C)]/(2\pi)$,
where $\zeta=x+\ii y$. For purposes of the approximation, let us restrict to
values of $\zeta$ for which these zeros are well-separated. The zeros coincide
if $P(0)=0$, which implies that $\zeta=\pm C$, so we have to stay away from these
values of $\zeta$.  Note that near $s=s_0$, we have $P(s)\approx2\pi\xi(s-s_0)$,
where $\xi=C\sinh(2\pi s_0)$.

Define $E(s)=\exp[-cs\bar{s}-z(s-\bar{s})/2]$,
where $c$ is a positive constant, and take $g=E(s-s_0)$. Then
(\ref{chainDeltaEqn}) is satisfied (to within our approximation) if and only if
$f= |\xi|g/\xi$ and $c=\pi|\xi|$.
In other words, one approximate solution of (\ref{chainDeltaEqn}) is
\[
 \left[\begin{array}{c} g\\f\end{array}\right]
 \approx\left[\begin{array}{c} \xi\\|\xi|\end{array}\right] E(s-s_0),
\]
which is strongly peaked at $s=s_0$.  The other (independent) solution
is peaked at $s=-s_0$, and is obtained similarly.  So we can take
\begin{equation} \label{chainPsi}
  \Psi \approx |\xi|^{-1/2}\left[\begin{array}{cc}
            \xi E(s-s_0) & -\xi E(s+s_0) \\
            |\xi| E(s-s_0) & |\xi| E(s+s_0)
         \end{array}\right],
\end{equation}
where the normalization factor (ensuring that $\int\Psi^{\dagger}\Psi=I$)
follows from
\[
  \int\!\!\int |E(r+\ii t)|^2\,dr\,dt \approx \frac{1}{2|\xi|}.
\]
Finally, from $\Psi$ we can compute the Higgs field, and we get
\begin{equation} \label{chainPhi}
  |\Phi|\approx|\Re(s_0)|=|\Re \cosh^{-1}(\zeta/C)|/(2\pi).
\end{equation}
Several things can immediately be deduced from~(\ref{chainPhi}):
\begin{itemize}
 \item $|\Phi|\sim[\log(2\rho/C)]/(2\pi)$ as $\rho\to\infty$, which agrees
       with the required boundary behaviour~(\ref{chainBC});
 \item $|\Phi|$ and ${\cal E}$ are independent of~$z$;
 \item $\Phi$ vanishes on the planar segment $-C < x < C$, $y=0$ (but bear in
       mind that our approximation is not guaranteed to hold near $x=\pm C)$;
 \item the energy density ${\cal E}$ is localized around the two lines
       $x=\pm C$, $y=0$ (again bearing in mind that this is exactly where
       the approximation is unclear).
\end{itemize}
Plotting $|\Phi|^2$ and
$|D\Phi|^2=\frac{1}{2}{\cal E}=\frac{1}{2}\nabla^2|\Phi|^2$ obtained both
from this approximation, and from a numerical implementation of the Nahm
transform, for $C=8$, yields the plots in Figure~1. The two upper subfigures
use the approximate solution, with $|D\Phi|^2$ truncated near $x=\pm C$, $y=0$;
while two lower subfigures
use the Higgs field obtained from the numerical Nahm transform.  The two
methods yield the same picture, except where one expects the approximation
to break down.
\begin{figure}[htb]
  \includegraphics[scale=0.8]{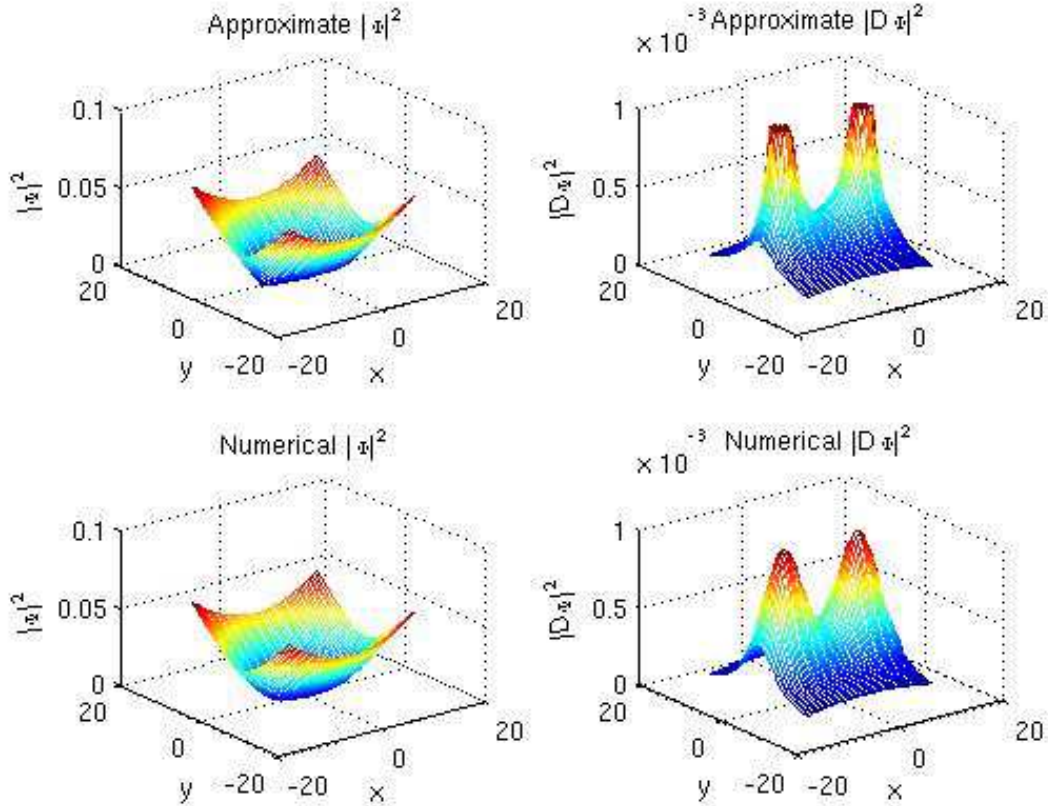}
  \caption{$|\Phi|^2$ and $|D\Phi|^2$ on the $xy$-plane, for $C=8$.
    The upper figures use the approximate solution, and the lower figures use
    a numerical solution. \label{fig1}}
\end{figure}

To summarize, the appearance of an infinite chain of 1-monopoles is as follows.
If $C$
(the ratio between the monopole size and the period) is small, then one has a
chain of small monopoles, each roughly spherical in shape, strung along a line
(in this case, the $z$-axis).  For large $C$, however, the energy density becomes
approximately constant in the $z$-direction, and is peaked along two lines
parallel to the $z$-axis, each a distance $C$ from it.  The numerical results
indicate that the zeros of the Higgs field $\Phi$ are located on the $z$-axis,
at $z=2\pi n$ for $n\in{\bf Z}$; but for large $C$, $\Phi$ is very close to
zero on the whole of the planar segment $-C < x < C$, $y=0$.


\section{Monopole Sheets}

By a monopole sheet we mean a solution of (\ref{Bog}) which is periodic in
two of the three dimensions (say the $x$ and $y$ directions), and satisfies
an appropriate boundary condition in the $z$-direction.  In other words,
the field lives on $\RR\times T^2$.  The general pattern for the Nahm
transform is that monopoles on $\RR^{3-l}\times T^l$ correspond to
solutions of (\ref{Bog}) on $\RR\times T^l$ which are independent of the
remaining $2-l$ coordinates.  The cases $l=0$ (monopoles on $\RR^3$
corresponding to solutions of the Nahm equations on $\RR$) and $l=1$
(\cite{CK01}, discussed in the
previous section) are well-established; but not much is known about the $l=2$
case. In view of the general pattern,
one would expect that the Nahm transform of a monopole on $\RR\times T^2$
will be another monopole on $\RR\times T^2$. It remains to be seen whether
or not this is the case in general (and under what circumstances), but we shall
see now that the simplest (abelian) example confirms this picture.

Consider the well-known homogeneous U(1) gauge field, with gauge potential
$A_j=\frac{1}{2}\ii B(-y,x,0)$. Here $B$ is a real constant, which represents
the magnetic flux density through the $xy$-plane.  With Higgs field
$\Phi=-\ii Bz$, we have a U(1) solution of the monopole equations (\ref{Bog}).
This field is doubly-periodic (up to gauge) in the $x$ and $y$ directions,
with periods
$\lambda_x$ and $\lambda_y$ respectively, provided we impose the Dirac
quantization condition $B\lambda_x\lambda_y=2\pi N$, with $N$ being an
integer.  Geometrically, $N$ is the first Chern class of the U(1) bundle
on $T^2$.  For simplicity in what follows, let us take
$\lambda_x=1=\lambda_y$ and $N=1$, so that $B=2\pi$. The Nahm transform of
the field involves the normalizable solutions of $\Delta\Psi=0$, where
\begin{equation} \label{sheetDelta}
  \Delta = \left[\begin{array}{cc}
            D_z-\ii\Phi-Z & 2D_s+\ii\overline{S} \\
            2D_{\bar s}+\ii S & -D_z-\ii\Phi-Z
         \end{array}\right].
\end{equation}
Here $s=x+\ii y$, $D_j=\pa_j+A_j$, $(X,Y,Z)$  are the dual coordinates ($X$
and $Y$ are periodic, with the dual period $2\pi$), and $S=X+\ii Y$.
The boundary conditions on $\Psi$ are
\begin{equation} \label{sheetBC}
  \left.
    \begin{array}{l}
      \Psi(x+1,y,z)=\Psi(x,y,z)\exp(-\ii\pi y) \\
      \Psi(x,y+1,z)=\Psi(x,y,z)\exp(\ii\pi x) \\
      \Psi(x,y,z)\to0 \quad\mbox{as $z\to\pm\infty$}
    \end{array}
  \right\}
\end{equation}
%
%
%
Putting in the homogeneous U(1) field described above gives the system
\begin{equation} \label{sheetDeltaEqn}
  \Delta \left[\begin{array}{c} g\\f\end{array}\right]
  = \left[\begin{array}{c} 
        g_z-(2\pi z+Z)g +2f_s+(\pi{\bar s}+\ii\overline{S})f \\
        2g_{\bar s}-(\pi s-\ii S)g -f_z-(2\pi z+Z)f
         \end{array}\right] = 0.
\end{equation}
A solution of (\ref{sheetDeltaEqn}), satisfying the required boundary
conditions, is $g=0$ and
\begin{equation} \label{sheetNahmSoln}
 f(x,y,z)=\Lambda\,\overline{\vartheta_3(\pi s-\ii S/2)}\,
   \exp\left[-\frac{1}{4\pi}(2\pi z+Z)^2 + (\overline{S}-\ii\pi{\bar s})y\right].
\end{equation}
Here $\vartheta_3$ is the theta-function
$\vartheta_3(\zeta)=1+2\sum_{k=1}^{\infty}\exp(-\pi k^2)\cos(2k\zeta)$,
and $\Lambda$ is a normalization constant determined by
$\int|f|^2\,d^3x=1$.  We can then compute the Nahm transform of $(\Phi,A_j)$:
these are U(1) fields on the $XYZ$-space, and are given by
\begin{eqnarray*}
  \widetilde{A}_j &=& \int\Psi^{\dagger}\frac{\pa}{\pa X^j}\Psi \,d^3x
                      = \frac{1}{2\pi}(0,-\ii X,0), \\
  \widetilde{\Phi} &=& -\ii\int z\Psi^{\dagger}\Psi \,d^3x
                      = \frac{\ii Z}{2\pi}.
\end{eqnarray*}
This is essentially the same solution as we started with (except that the
periods are dual to the original ones).  So this U(1) monopole sheet is
``self-reciprocal'' \cite{CG84} under the Nahm transform.


What about non-abelian monopole sheets?  By analogy with the abelian case,
we expect the boundary condition in $z$ to be that $\Phi$ is linear in $z$,
and $|D\Phi|$ tends to a positive constant, as $z\to\infty$.
In \cite{L99}, it was argued that the embedding of the U(1) example into SU(2)
may legitimately be thought of as an SU(2) monopole sheet.  Part of the
argument came from looking at the normalizable zero-modes of the embedding.
The calculations in \cite{L99} did not impose periodicity in the $xy$-plane,
and the version below is a variant which does.

Let us write $\Phi=\Phi^a\,\sigma^a/(2\ii)$,
$A_j = A_j^a\,\sigma^a/(2\ii)$,
where $\sigma^a$ denote the Pauli matrices.  The embedded solution is
\[
  \widehat{\Phi} =-2\pi z \frac{\sigma^3}{2\ii}, \quad
    \widehat{A}_j =\pi(-y,x,0)\frac{\sigma^3}{2\ii}.
\]
We consider a perturbation
\begin{equation} \label{sheetFluct0}
  \Phi=\widehat{\Phi}+\phi, \quad A_j=\widehat{A}_j+a_j,
\end{equation}
where $\phi$ and $a_j$ are infinitesimal.
We can take $\phi^3=0=a_j^3$, since we are only interested
in ``non-abelian'' fluctuations.  Writing $W_j=a^1_j+\ii a^2_j$ and
$W_4=\phi^1+\ii\phi^2$, and imposing the monopole equations
(\ref{Bog}) together with a gauge condition $D_j a_j+[\Phi,\phi]=0$,
gives the system
\begin{equation} \label{sheetFluct1}
  \Delta_0 \left[\begin{array}{c}
     W_1-\ii W_2 \\ -W_3+\ii W_4\end{array}\right]
  = 0 =
  \Delta_0 \left[\begin{array}{c}
     W_3+\ii W_4 \\ W_1+\ii W_2\end{array}\right],
\end{equation}
where $\Delta_0$ is the operator (\ref{sheetDelta}) with
$Z=S=\overline{S}=0$.  The same boundary conditions (\ref{sheetBC})
as before apply here
as well, in order for the perturbation to be normalizable and
doubly-periodic.  So a solution of (\ref{sheetFluct1}) is given by
\begin{equation} \label{sheetFluct2}
W_1=\ii W_2=\alpha f_0, \quad W_4=\ii W_3=\beta f_0,
\end{equation}
where $\alpha$ and $\beta$ are complex constants,
and $f_0$ is the function (\ref{sheetNahmSoln}) with $Z=S=\overline{S}=0$.
This suggests \cite{L99} that the ``abelian'' monopole sheet belongs to a
four-parameter family of doubly-periodic SU(2) monopole sheets.  However, it
remains to be shown whether these actually exist --- in other words, whether
the zero-modes (\ref{sheetFluct2}) correspond to actual solutions.

To see what these solutions may look like, however, we can plot the
perturbed fields
(\ref{sheetFluct0}).  Clearly the perturbations are concentrated on
the plane $z=0$.  In Figure 2, the quantities $|\Phi|^2$ and $|D\Phi|^2$
are plotted on $z=0$, for $0\leq x\leq2$ and $0\leq y\leq2$ (covering four
fundamental cells).  The field $\Phi$ is
obtained from (\ref{sheetFluct0}) and (\ref{sheetFluct2}),
with coefficient $\beta=0.5$.  The doubly-periodic nature of the field
is evident.  The unperturbed Higgs field $\widehat{\Phi}$ is identically
zero on $z=0$, whereas the perturbed field $\Phi$ has exactly one zero
in each cell; $\Phi$ is non-zero for $z\neq0$, and grows linearly with $z$.
Similarly, the energy density $|D\Phi|^2$ takes the constant
value $B^2/4=\pi^2$ in the unperturbed case, whereas the perturbed version
is non-constant near $z=0$ and is peaked where $\Phi$ has its zero.  
\begin{figure}[htb]
\begin{center}
\subfigure[$|\Phi|^2$ on $z=0$]{
\includegraphics[scale=0.35]{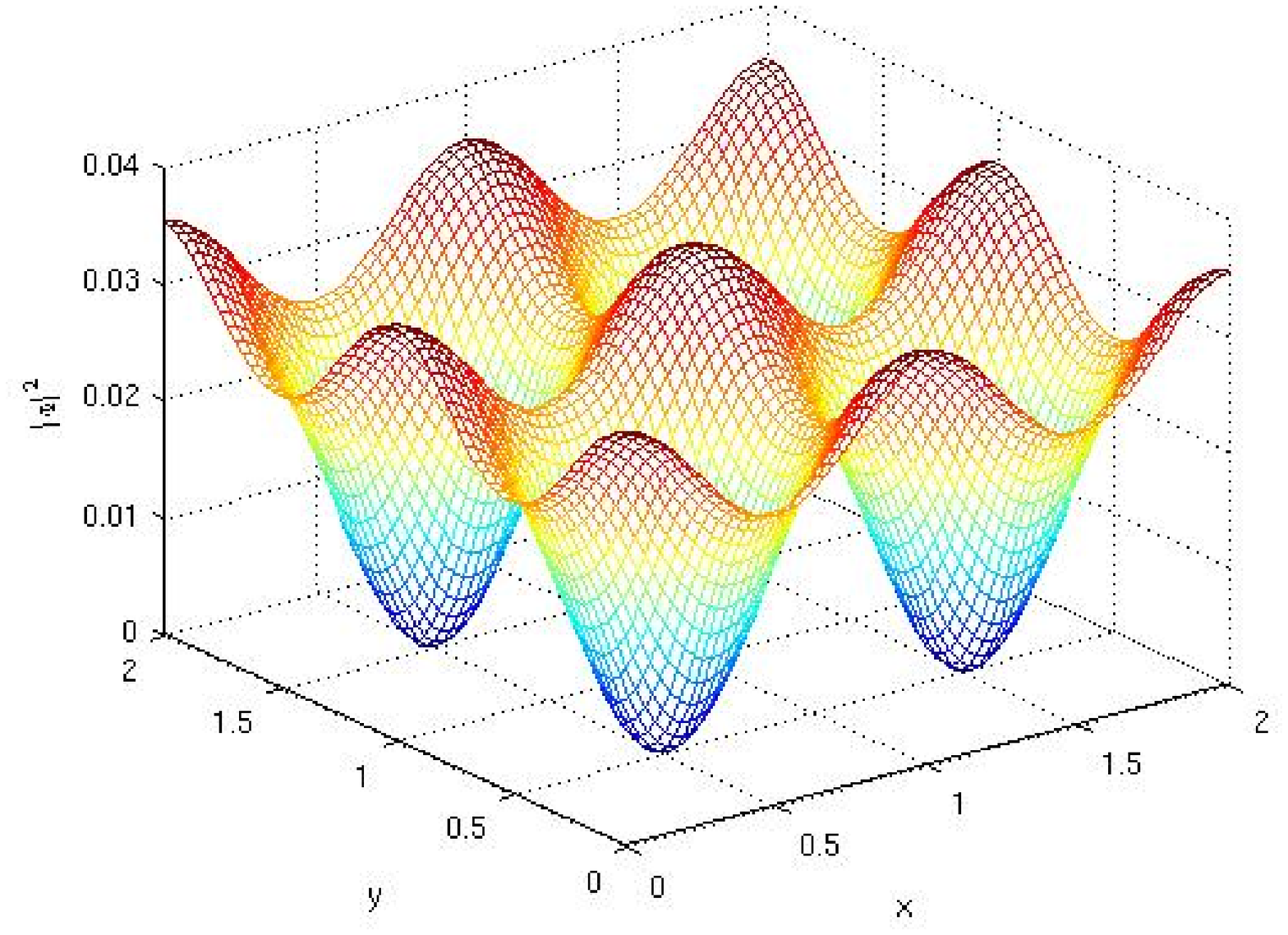}
}
\quad
\subfigure[$|D\Phi|^2-\pi^2$ on $z=0$]{
\includegraphics[scale=0.35]{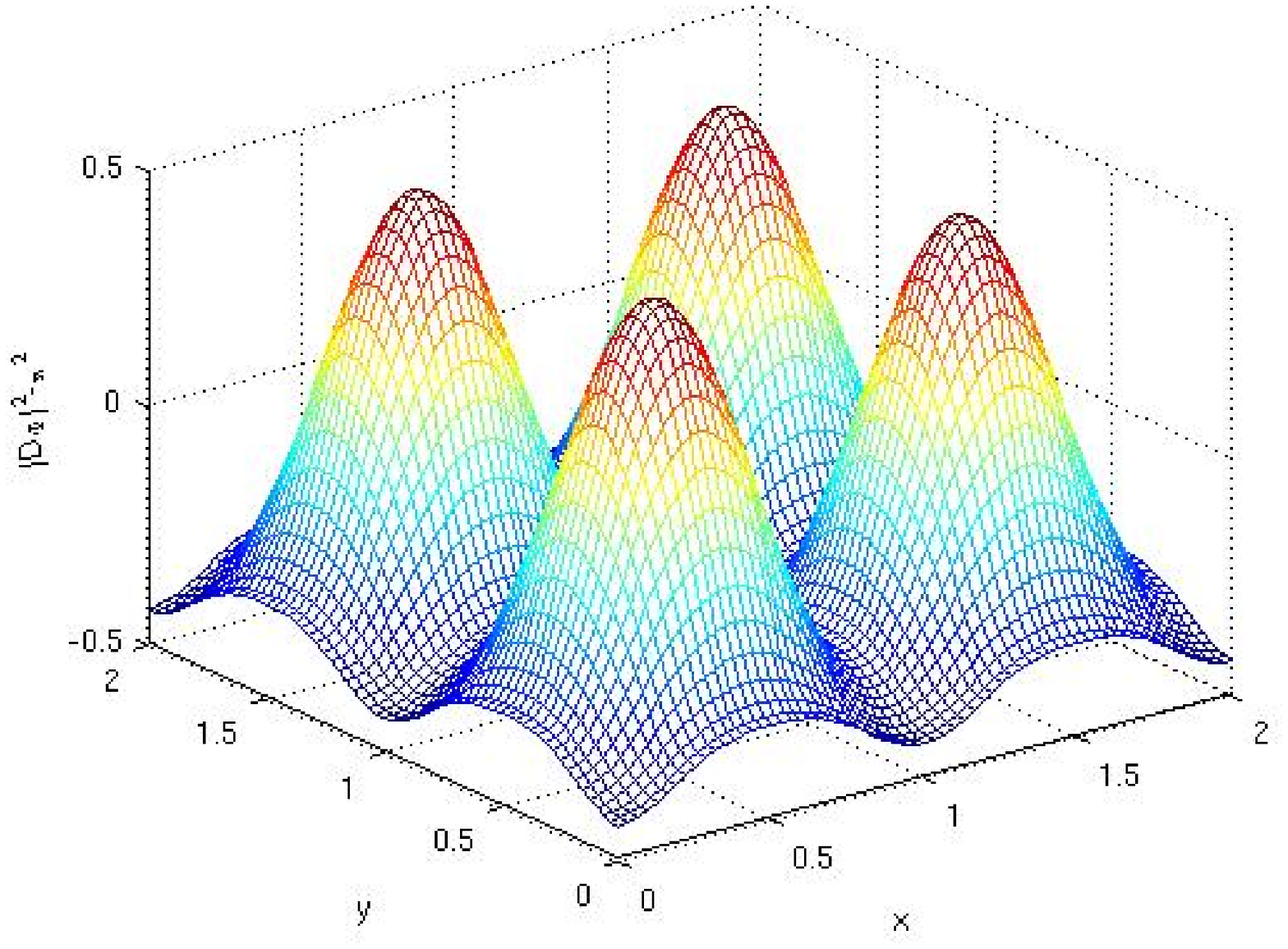}
}
\caption{SU(2) monopole sheet: perturbation of homogeneous solution.
          \label{fig2}}
\end{center}
\end{figure}

Clearly much analysis remains to be done in this case, to confirm that
solutions exist, understand their moduli space, and work out the
details of the Nahm transform. Work in this direction is currently under way.


\bigskip\noindent{\bf Acknowledgments.}
This work was supported by a research grant from the UK Engineering and
Physical Sciences Research Council, and by the grant ``Classical Lattice
Field Theory'' from the UK Particle Physics and Astronomy Research Council.


\end{document}